# Enhancing the STIX Representation of MITRE ATT&CK for Group Filtering and Technique Prioritization

Mateusz Zych and Vasileios Mavroeidis

University of Oslo, Oslo, Norway

mateusdz@ifi.uio.no

vasileim@ifi.uio.no

**Abstract:** In this paper, we enhance the machine-readable representation of the ATT&CK Groups knowledge base provided by MITRE in STIX 2.1 format to make available and queryable additional types of contextual information. Such information includes the motivations of activity groups, the countries they have originated from, and the sectors and countries they have targeted. We demonstrate how to utilize the enhanced model to construct intelligible queries to filter activity groups of interest and retrieve relevant tactical intelligence.

## 1. Introduction

The exponential increase in cyberattacks and sophisticated attack behavior push organizations to continuously invest in strengthening their cybersecurity posture (Brown & Lee 2021). Defenders to better respond to the current cyber threat landscape, improve their threat situational awareness, and stay resilient against cybersecurity threats have recognized the need to generate and utilize cyber threat intelligence and collaborate through information exchange to use others' experiences as their own organization's defense (Johnson et al. 2016).

There is a common understanding that for detection purposes, the value of atomic indicators like hash values, IP addresses, domain names, and host and network artifacts can vary due to their possible short lifespan and their vast amount that often leads to indicator and alert fatigue. David Bianco depicts that through a "pyramid of pain" (Bianco 2014). Bianco's pyramid of pain is shown in Figure 1 and represents the pain an adversary will suffer when a defender has denied the adversary the use of those indicators. At the apex of the pyramid of pain are Tactics, Techniques, and Procedures (TTPs) as a single grouping. When defenders detect and respond at the TTP level, they operate directly on adversary behavior. Detection at the TTP level can identify malicious activity that may not rely on prior knowledge of adversary tools and atomic indicators and can have a lasting impact on the attackers as they would have to change nearly every aspect of how they operate to avoid detection (Lee & Brown 2021).

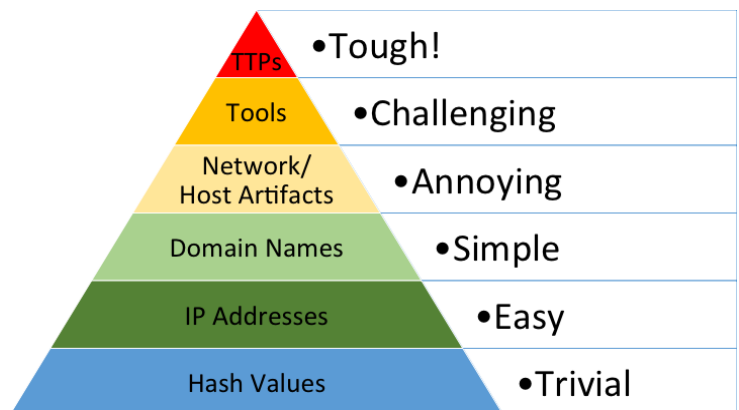

**Figure 1:** The Pyramid of Pain (Bianco 2014).

A globally-accessible effort that curates knowledge concerning adversary behavior is ATT&CK (Adversarial Tactics, Techniques, and Common Knowledge), led by MITRE and supported by the broader cybersecurity community via contributions. ATT&CK focuses on how adversaries compromise and operate within computer information networks. ATT&CK fundamentally is a set of taxonomies that provides defenders with a common language to map and communicate their findings pertinent to adversary behavior. It can support different areas of cyberspace defense, such as adversary emulation, behavioral analytics, cyber threat intelligence enrichment, defense gap assessment, red teaming, and SOC maturity assessment (Alexander, Belisle & Steele 2020). Furthermore, in the same project family, MITRE maintains a knowledge base of activity groups[1] (referred to as Groups) and their technique use. For programmatic utilization, the entire ATT&CK knowledge base can be accessed through STIX interfaces that MITRE provides on their GitHub[2].

This paper deals with enhancing the existing representation approach (model) of ATT&CK Groups in STIX 2.1 as a means of improving the amount and types of contextual information made available in a fully structured way. In particular, we focus on information that currently lies dormant in a semi-structured way in the description text field (see Table 1) of activity groups, such as group motivations, the countries they have originated from, and sectors and countries they have targeted. Using the proposed enhanced STIX 2.1 model, we demonstrate how to construct intelligible queries to filter activity groups of interest and retrieve relevant tactical intelligence.

The rest of the paper is structured as follows. Section 2 introduces the reader to ATT&CK Groups and discusses how the knowledge base is currently represented in STIX 2.1. Section 3 presents our proposed STIX 2.1 model that extends the current representation of the Groups knowledge base and discusses the set of new objects we utilize to make new types of information available. Section 4 demonstrates a use case where we use the proposed model to construct intelligible queries to filter activity groups of interest and then prioritize ATT&CK techniques that we should establish or validate controls against. Finally, section 5 concludes the paper.

## 2. ATT&CK Groups and Their Structure

Groups[3] are sets of related intrusion activity that are tracked by a common name in the security community and are aimed to collect ATT&CK techniques and software that have been reported to use. Some groups have multiple names associated with similar activities due to various organizations tracking similar activities by different names. Table 1 shows the set of properties (characterized as data items in the first column of Table 1) that formulate a single group. Briefly, a property of type "tag" points to an informational reference of a group, a property of type "field" includes information as free text, and a property of type "relationship" references an object that is directly related to the group such as ATT&CK techniques and software. Of particular importance for this research is the "description" property that, in many cases, carries information about a group's motivation, the country it originates from, as well as sectors and countries that have been targeted. As we will see in the following sections, we aim to make such information directly accessible and, consequently, queryable by extending the STIX 2.1 representation of the knowledge base.

---

[1] https://attack.mitre.org/groups/
[2] https://github.com/mitre-attack/attack-stix-data
[3] https://attack.mitre.org/groups/

**Table 1:** ATT&CK Group model (Strom et al. 2020).

| Data Item | Type | Description |
| --- | --- | --- |
| **Name** | Field | The name of the adversary group. |
| **ID** | Tag | Unique identifier for the group within the knowledgebase. Format: G####. |
| **Associated Groups** | Tag | Names that have overlapping references to a group entry and may refer to the same or similar group in threat intelligence reporting. |
| **Version** | Field | Version of the group in the format of MAJOR.MINOR. |
| **Contributors** | Tag | List of non-MITRE contributors (individual and/or organization) from first to most recent that contributed information on, about, or supporting the development of a group profile. |
| **Description** | Field | A description of the group based on public threat reporting. It may contain dates of activity, suspected attribution details, targeted industries, and notable events that are attributed to the group's activities. |
| **Associated Group Descriptions** | Field | Section that can be used to describe the associated group names with references to the report used to tie the associated group to the primary group name. |
| **Techniques / Sub-Techniques Used** | Relationship / Field | List of (sub-)techniques that are used by the group with a field to describe details on how the technique is used. This represents the group's procedure (in the context of TTPs) for using a technique. Each technique should include a reference. |
| **Software** | Relationship / Field | List of software that the group has been reported to use with a field to describe details on how the software is used. |

## 2.1 MITRE's STIX 2.1 Representation of ATT&CK Groups

The STIX standard developed by the OASIS Cyber Threat Intelligence Technical Committee is a common language and serialization format used to represent and exchange cyber threat intelligence (Jordan, Piazza & Darley 2021). MITRE provides access to the ATT&CK knowledge base in STIX 2.0 and STIX 2.1. For the purpose of this research, we processed and manipulated version[4] 9 of the knowledge base in STIX 2.1. It is worth noting that the official web interface that allows users to interact with ATT&CK is built from the STIX data.

Figure 2 illustrates the set of STIX 2.1 object and relationship types leveraged to define groups that altogether comprise the ATT&CK Groups knowledge base. An "intrusion set" object (activity group or Group) connects with "attack pattern" objects (techniques/sub-techniques), "malware" objects (software), and "tool" objects (software).

---

[4] https://github.com/mitre/cti/tree/eb1b9385d44340ce867a77358c5f5aaed666e54c/enterprise-attack/intrusion-set

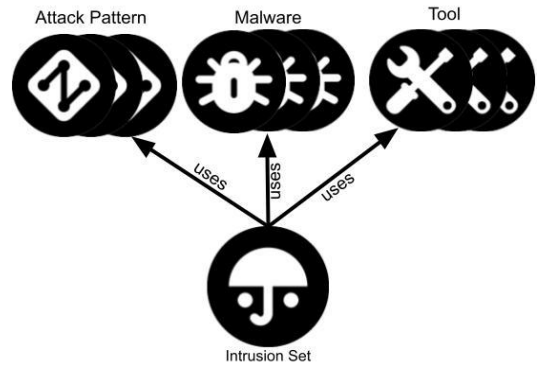

**Figure 2:** High-level STIX 2.1 representation of the Groups knowledge base.

The intrusion set object type is used to represent activity groups. An intrusion set is a grouped set of adversarial behaviors and resources with common properties that is believed to be orchestrated by a single organization (Jordan, Piazza & Darley 2021). An attack pattern object describes the ATT&CK techniques and sub-techniques used in the intrusion set. A malware object represents commercial, custom closed source, or open-source software used for malicious purposes by adversaries (MITRE ATT&CK 2022). A tool object represents commercial, open-source, built-in, or publicly available software that could be used by a defender, pentester, red teamer, or adversary, and can include both software that generally is not found on an enterprise system as well as software generally available as part of an operating system that is already present in an environment (MITRE ATT&CK 2022).

# 3. A Proposed Enhanced STIX 2.1 Representation for the ATT&CK Groups Knowledge Base

As previously mentioned, this research proposes enhancements to MITRE's STIX 2.1 representation of the ATT&CK Groups knowledge base. We introduce additional types of contextual information that enable us to construct more intelligible queries and exploit the knowledge base in a more granular manner. Our proposed enhancements rely on information already part of the knowledge base but represented in a semi-structured form. In particular, information from the descriptions of activity groups, which in many cases, identify group motivations, the countries they originate from, as well as sectors and countries they have targeted. Example 1 presents such a case.

Example 1: Description of APT29 from the MITRE ATT&CK (version 9) Groups knowledge base[5].

> "APT29 is a threat group that has been attributed to Russia's Foreign Intelligence Service (SVR). They have operated since at least 2008, often targeting government networks in Europe and NATO member countries, research institutes, and think tanks. APT29 reportedly compromised the Democratic National Committee starting in the summer of 2015.
> In April 2021, the US and UK governments attributed the SolarWinds supply chain compromise cyber operation to the SVR; public statements included citations to APT29, Cozy Bear, and The Dukes. Victims of this campaign included government, consulting, technology, telecom, and other organizations in North America, Europe, Asia, and the Middle East. Industry reporting referred to the actors involved in this campaign as UNC2452, NOBELIUM, StellarParticle, and Dark Halo."

---

[5] https://attack.mitre.org/groups/G0016/

We extract the identified information and represent it in a fully structured manner using a set of distinct STIX 2.1 objects. In particular, the "location" object type is utilized to represent the suspected or confirmed country of origin of an activity group as well as targeted countries and regions. The "identity" object type is used to represent targeted sectors. Moreover, when feasible, the STIX 2.1 "intrusion set" objects are enriched with primary and secondary motivations (using the available primary and secondary motivations properties of the intrusion set object type). Figure 3 depicts the enhanced STIX 2.1 representation of the Groups knowledge base (version 9).

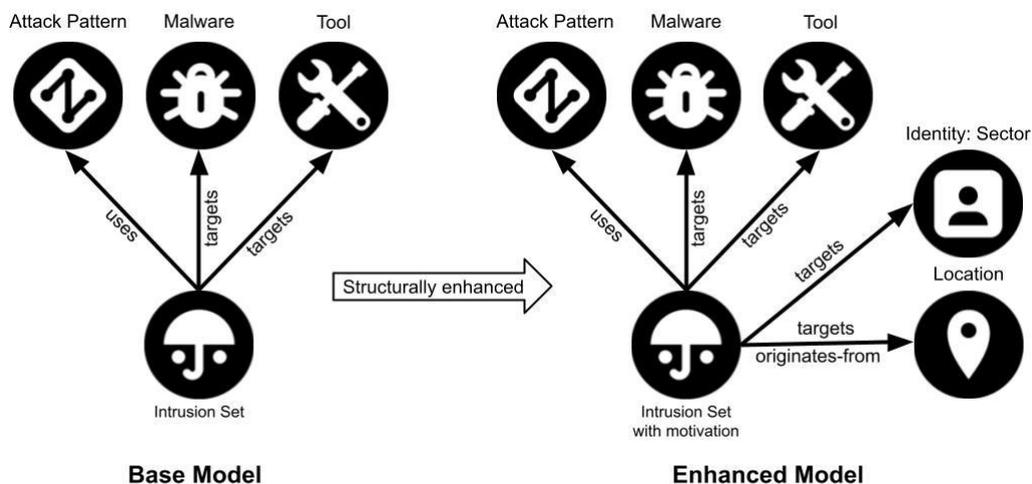

**Figure 3:** Enhanced STIX 2.1 representation of the ATT&CK Groups knowledge base.

Similarly, Figure 4, based on our proposed model, depicts an enhanced STIX 2.1 representation of APT29 with the contextual prose presented in Example 1 transformed into structured queryable data elements. We have made the entire enhanced Groups knowledge base available on GitHub[6].

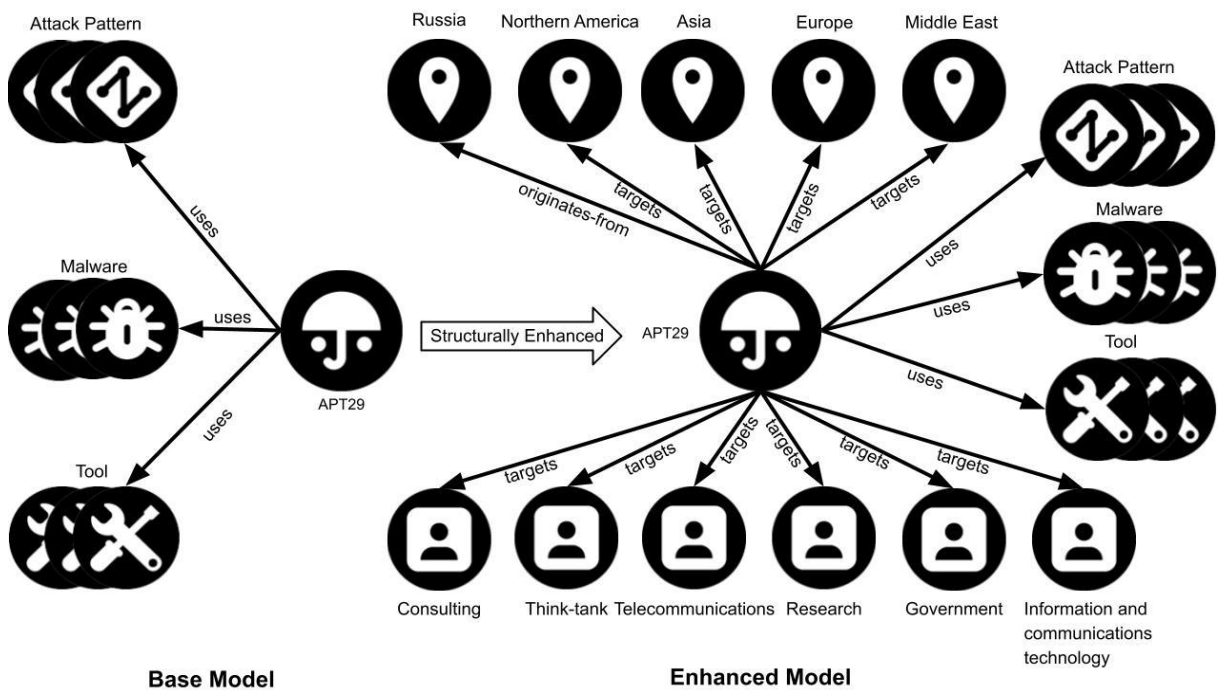

**Figure 4:** Graphical representation of APT29 using our enhanced model.

---

[6] https://github.com/fovea-research/SAG

# 4. Utilizing the Enhanced STIX 2.1 Groups Knowledge Base for Group Filtering and Technique Prioritization

This section demonstrates how to use the new types of contextual information introduced in the enhanced STIX 2.1 representation of the Groups knowledge base to filter activity groups of interest.

It is out of the scope of this paper to suggest specific data query- or programming languages to interact with the knowledge base. However, any programming language that can interact with JSON (JavaScript Object Notation) should be sufficient. It is also important to remark that the accuracy of the results retrieved might be questionable due to the use of non-standardized vocabularies in providing context that often leads to semantic ambiguity. For example, the number of sectors and sectors considered critical infrastructure might be dissimilar in different countries.

Listing 1 presents an SQL-based query that aims to retrieve activity clusters that are believed to originate from the Russian Federation and have targeted establishments within the Government sector in the United States. The results indicate three activity groups to be of relevance.

```
SELECT * FROM GroupsKnowledgeBase
WHERE OriginatesFrom == "Russian Federation"
AND TargetSector == "Government"
AND TargetCountry == "United States"

Results:
APT28, APT29, Dragonfly 2.0
```

**Listing 1:** SQL query leveraging our model's contextual information types to retrieve activity groups of interest.

Awareness of the techniques and their frequency among relevant activity groups allows organizations to support decision-making to protect their assets. From that point, defenders can leverage the associated information for their use cases, such as prioritizing adversary techniques to conduct defense gap- or SOC (security operations center) maturity assessments. We retrieve the ATT&CK technique use of the groups and provide it as input to MITRE's ATT&CK Navigator[7] to identify and explore any commonalities. The technique overlap is presented in Figure 5. The light red color techniques do not overlap among the three groups. The red color techniques indicate use by two groups, and the dark red color techniques indicate use by all three groups.

# 5. Conclusion

In this paper, we extended (enhanced) the STIX 2.1 model of ATT&CK Groups with additional types of contextual information. Such information includes the groups' motivations, countries of origin, and targeted sectors and countries. We demonstrated how to use the proposed model to filter activity groups of interest and consequently retrieve relevant tactical intelligence (behaviors/techniques). We used the ATT&CK techniques we extracted to identify and explore commonalities in technique use among different activity groups.

---

[7] https://mitre-attack.github.io/attack-navigator/

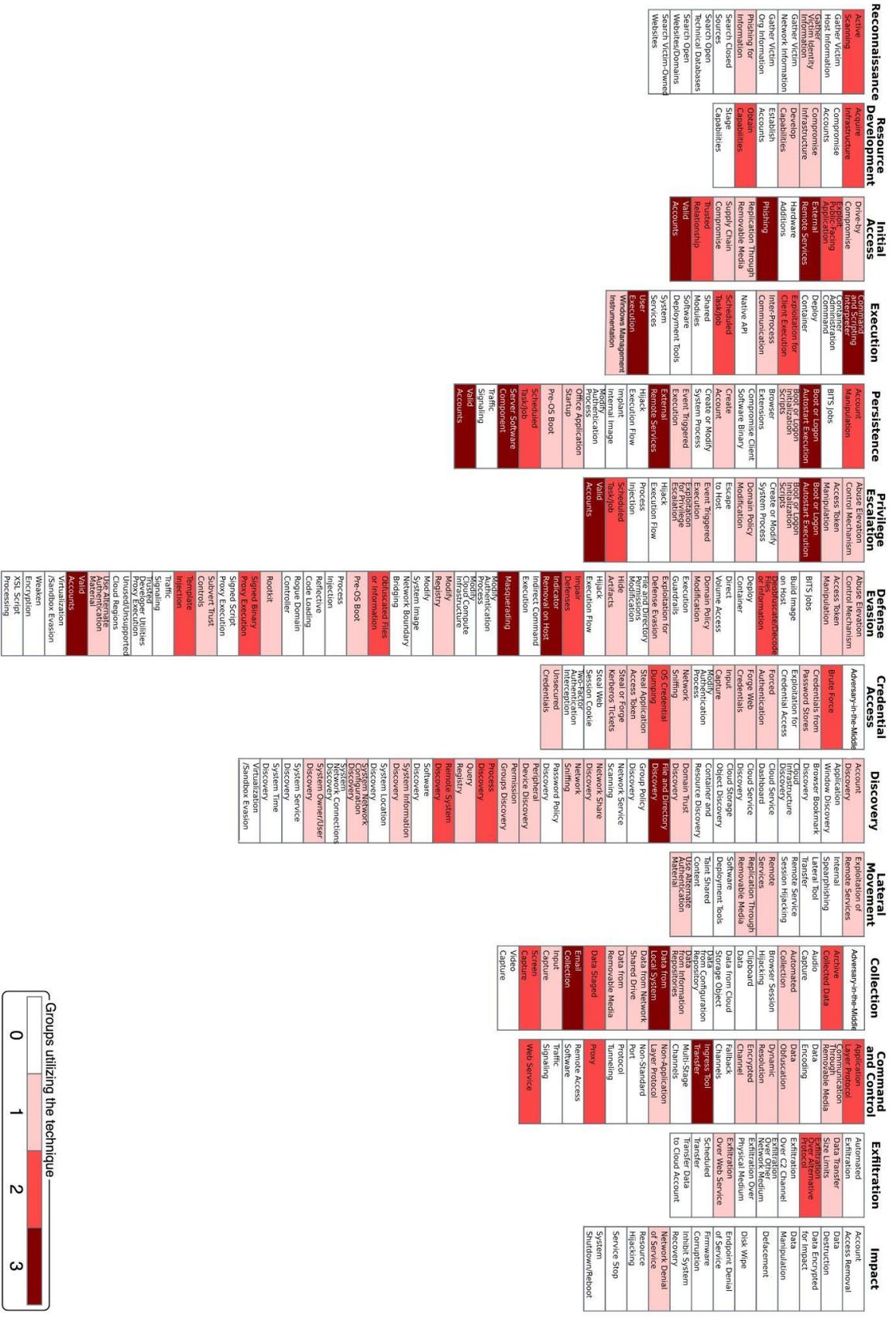

**Figure 5:** Heatmap showing commonalities in technique use among APT28, APT29, and Dragonfly 2.0.

## Acknowledgments

This research has received funding from the Research Council of Norway (forskningsrådet) under Grant Agreement No. 303585. In addition, this research work was supported by the European Health and Digital Executive Agency (HaDEA) under Grant Agreement No. INEA/CEF/ICT/A2020/2373266. The authors would like to express their appreciation to EclecticIQ for designing and making available icons to visualize the STIX language[8].

---

[8] https://github.com/eclecticiq/stix-icons